\documentclass[aps,prb,superscriptaddress,reprint]{revtex4-1}

\usepackage{graphicx}
\usepackage{subfigure}
\usepackage{docs}
\usepackage{bm}
\usepackage{amssymb}
\usepackage{amsmath}
\usepackage[colorlinks=true,linkcolor=blue]{hyperref}
\expandafter\ifx\csname package@font\endcsname\relax\else
 \expandafter\expandafter
 \expandafter\usepackage
 \expandafter\expandafter
 \expandafter{\csname package@font\endcsname}%
\fi
\begin{document}

\widetext
\title{Valence band-anticrossing in GaP$_{1-x}$Bi$_{x}$ dilute bismide alloys:\\ giant bowing of the band gap and spin-orbit splitting energy}



\author{Zoe L.~Bushell}
\affiliation{\mbox{Advanced Technology Institute \& Department of Physics, University of Surrey, Guildford, GU2 7XH, U.K.}}

\author{Christopher A.~Broderick}
\affiliation{\mbox{Tyndall National Institute, Lee Maltings, Dyke Parade, Cork T12 R5CP, Ireland}}
\affiliation{\mbox{Department of Electrical and Electronic Engineering, University of Bristol, Bristol BS8 1UB, U.K.}}

\author{Lukas Nattermann}
\affiliation{\mbox{Materials Science Center \& Faculty of Physics, Philipps-Universit{\"a}t Marburg, 35032 Marburg, Germany}}

\author{Rita~M.~Joseph}
\affiliation{\mbox{Advanced Technology Institute \& Department of Physics, University of Surrey, Guildford, GU2 7XH, U.K.}}

\author{Joseph L.~Keddie}
\affiliation{\mbox{Advanced Technology Institute \& Department of Physics, University of Surrey, Guildford, GU2 7XH, U.K.}}

\author{Judy M.~Rorison}
\affiliation{\mbox{Department of Electrical and Electronic Engineering, University of Bristol, Bristol BS8 1UB, U.K.}}

\author{Kerstin Volz}
\affiliation{\mbox{Materials Science Center \& Faculty of Physics, Philipps-Universit{\"a}t Marburg, 35032 Marburg, Germany}}

\author{Stephen J.~Sweeney}
\email{s.sweeney@surrey.ac.uk}
\affiliation{\mbox{Advanced Technology Institute \& Department of Physics, University of Surrey, Guildford, GU2 7XH, U.K.}}

\date{\today}



\begin{abstract}

Using spectroscopic ellipsometry measurements on GaP$_{1-x}$Bi$_{x}$/GaP epitaxial layers up to $x = 3.7$\% we observe a giant bowing of the direct band gap ($E_{g}^{\Gamma}$) and valence band spin-orbit splitting energy ($\Delta_{\scalebox{0.6}{\textrm{SO}}}$). $E_{g}^{\Gamma}$ ($\Delta_{\scalebox{0.6}{\textrm{SO}}}$) is measured to decrease (increase) by approximately 200 meV (240 meV) with the incorporation of 1\% Bi, corresponding to a greater than fourfold increase in $\Delta_{\scalebox{0.6}{\textrm{SO}}}$ in going from GaP to GaP$_{0.99}$Bi$_{0.01}$. The evolution of $E_{g}^{\Gamma}$ and $\Delta_{\scalebox{0.6}{\textrm{SO}}}$ with $x$ is characterised by strong, composition-dependent bowing. We demonstrate that a simple valence band-anticrossing model, parametrised directly from atomistic supercell calculations, quantitatively describes the measured evolution of $E_{g}^{\Gamma}$ and $\Delta_{\scalebox{0.6}{\textrm{SO}}}$ with $x$. In contrast to the well-studied GaAs$_{1-x}$Bi$_{x}$ alloy, in GaP$_{1-x}$Bi$_{x}$ substitutional Bi creates localised impurity states lying energetically within the GaP host matrix band gap. This leads to the emergence of an optically active band of Bi-hybridised states, accounting for the overall large bowing of $E_{g}^{\Gamma}$ and $\Delta_{\scalebox{0.6}{\textrm{SO}}}$ and in particular for the giant bowing observed for $x \lesssim 1$\%. Our analysis provides insight into the action of Bi as an isovalent impurity, and constitutes the first detailed experimental and theoretical analysis of the GaP$_{1-x}$Bi$_{x}$ alloy band structure.

\end{abstract}

\maketitle


Highly-mismatched III-V semiconductor alloys containing dilute concentrations of Bi have attracted significant attention in recent years \cite{Li_book_2013} since their unique electronic properties open up a range of possibilities for practical applications for near- and mid-infrared photonic devices, such as semiconductor lasers, \cite{Sweeney_ICTON_2011,Broderick_SST_2012,Marko_APL_2012,Sweeney_JAP_2013,Ludewig_APL_2013,Marko_JPDAP_2014,Fuyuki_APE_2014,Butkute_EL_2014,Marko_SR_2016,Kim_SST_2017,Wu_ACSP_2017,Marko_IEEEJSTQE_2017,Jin_JAP_2013,Broderick_NUSOD_2016} photovoltaics, \cite{Thomas_SST_2015,Richards_IEEEPVSC_2016}  spintronics, \cite{Mazzucato_APL_2013,Pursley_APL_2013,Simmons_APL_2015} photodiodes, \cite{Lee_APL_1997,Hunter_IEEEPTL_2012,Sandall_APL_2014,Gu_APL_2016} and thermoelectrics. \cite{Dongmo_JAP_2012} Research on dilute bismide alloys has primarily focused to date on GaAs$_{1-x}$Bi$_{x}$, where incorporation of Bi brings about a strong reduction of the direct $\Gamma$-point band gap ($E_{g}^{\Gamma}$) -- by up to 90 meV per \% Bi at low Bi compositions $x$ \cite{Francoeur_APL_2003,Yoshida_JJAP_2003,Alberi_PRB_2007,Usman_PRB_2011,Batool_JAP_2012} -- characterised by strong, composition-dependent bowing. \cite{Alberi_PRB_2007,Broderick_SST_2013} This unusual behaviour derives from the large differences in size (covalent radius) and chemical properties (electronegativity) between As and Bi: Bi, being significantly larger and more electropositive than As, acts as an isovalent impurity which primarily impacts and strongly perturbs the valence band (VB) structure. \cite{Zhang_PRB_2005,Deng_PRB_2010,Usman_PRB_2011} This is in contrast to dilute nitride alloys, in which small electronegative nitrogen (N) atoms strongly perturb the conduction band (CB) structure in GaN$_{x}$As$_{1-x}$ and related alloys. \cite{Shan_PRL_1999,Kondow_IEEEJSTQE_1997,Kent_PRB_2001,Reilly_SST_2009} Additionally Bi, being the largest stable group-V element, has strong relativistic (spin-orbit coupling) effects. \cite{Carrier_PRB_2004} As such, the reduction of $E_{g}^{\Gamma}$ in (In)GaAs$_{1-x}$Bi$_{x}$ is accompanied by a strong increase in the VB spin-orbit splitting energy ($\Delta_{\scalebox{0.6}{\textrm{SO}}}$). \cite{Fluegel_PRL_2006,Usman_PRB_2011,Batool_JAP_2012}

Epitaxial growth of GaP$_{1-x}$Bi$_{x}$ alloys, via molecular beam epitaxy \cite{Christian_APE_2015,Christian_JJAP_2016} and metal-organic vapour phase epitaxy \cite{Nattermann_JCG_2017_1} (MOVPE), has only recently been attempted. Here, we present the first detailed analysis of the GaP$_{1-x}$Bi$_{x}$ electronic band structure. Early experiments on impurities in GaP can be traced back to the advent of semiconductors with the initial experiments of Trumbore et al.\cite{Trumbore_APL_1966} revealing that Bi dopants generate bound localised impurity states in GaP, i.e.~Bi-related localised impurity states lying energetically within the GaP host matrix band gap. However, there is little further data iavailable regarding the GaP$_{1-x}$Bi$_{x}$ band structure. In this work, we explicitly verify that the evolution of the main features of the GaP$_{1-x}$Bi$_{x}$ VB structure with Bi composition, $x$, can be understood in a straightforward manner in terms of a Bi composition-dependent valence band-anticrossing (VBAC) interaction between the extended states of the GaP VB edge and highly localised bound impurity states associated with substitutional Bi impurities. Our measurements reveal giant bowing of $E_{g}^{\Gamma}$ and $\Delta_{\scalebox{0.6}{\textrm{SO}}}$: $E_{g}^{\Gamma}$ ($\Delta_{\scalebox{0.6}{\textrm{SO}}}$) decreases (increases) by $\approx 200$ meV (240 meV) when 1\% Bi is incorporated substitutionally in GaP. Comparison between theory and experiment highlights the emergence of an impurity band of primarily Bi-derived states, lying energetically within the GaP band gap but close in energy to the unperturbed GaP VB edge. The VBAC interaction leads to these states acquiring an admixture of localised (Bi) and extended $\Gamma_{8v}$ VB edge (Bloch) character, enabling optical coupling to the comparatively unperturbed  $\Gamma_{6c}$ CB states.


\begin{figure*}[t!]
     \centering
     \includegraphics[width=1.00\textwidth]{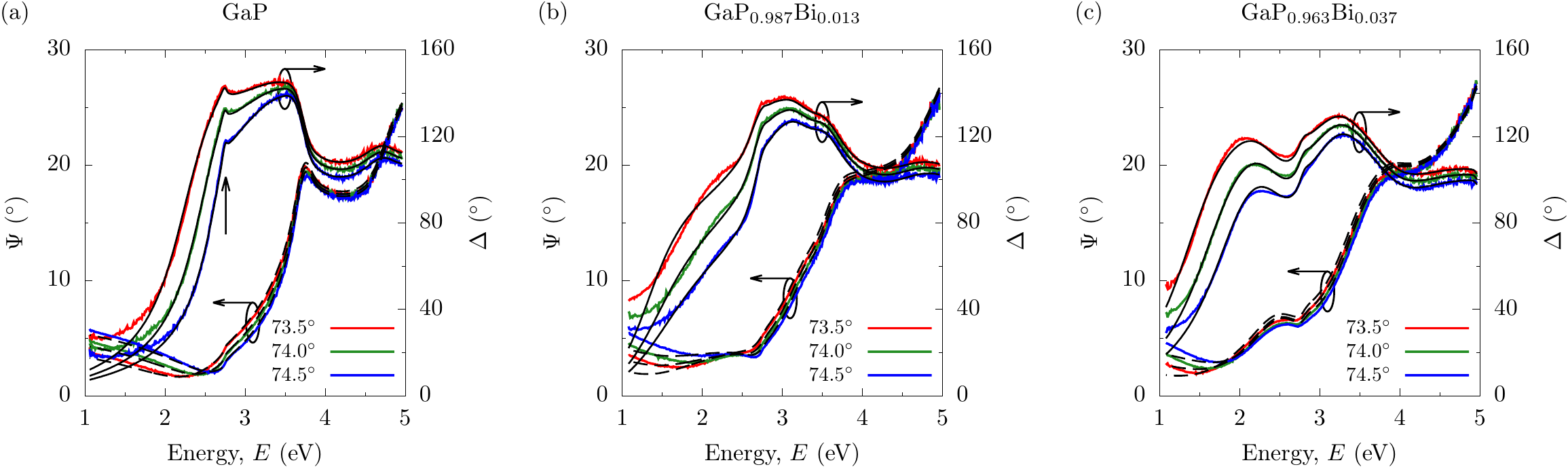}
     \caption{Measured SE spectra for the MOVPE-grown (a) GaP (Bi-free, $x = 0$), (b) GaP$_{0.987}$Bi$_{0.013}$ ($x = 1.3$\%), and (c) GaP$_{0.963}$Bi$_{0.037}$ ($x = 3.7$\%) samples described in the text and in Ref.~\onlinecite{Nattermann_JCG_2017_1}. Solid red, green and blue lines respectively denote data measured for incident beam angles of 73.5$^{\circ}$, 74.0$^{\circ}$ and 74.5$^{\circ}$. Solid (dashed) black lines show the fits to the measured $\Delta$ ($\Psi$) spectra; the SE model and associated fitting prodecure are outlined in the text.}
     \label{fig:ellipsometry_experiment}
\end{figure*}


Spectroscopic ellipsometry (SE) was used to study bulk-like GaP$_{1-x}$Bi$_{x}$/GaP epitaxial layers containing up to 3.7\% Bi. The samples upon which our analysis was performed were grown on (001)-oriented GaP by MOVPE. Full details of the sample growth and characterisation can be found in Ref.~\onlinecite{Nattermann_JCG_2017_1}. The SE measurements were performed at room temperature using a J.~A.~Woollam Co.~variable angle spectroscopic ellipsometer system. Three incident beam angles were used to generate sufficient data to provide confidence in modelling fits to the measured spectra. Angles of 73.5$^{\circ}$, 74.0$^{\circ}$ and 74.5$^{\circ}$ were chosen since they are close to the pseudo-Brewster angles of the samples under investigation, thereby ensuring that the phase change on reflection $\Delta$ remained close to 90$^{\circ}$. The sensitivity of the rotating analyser ellipsometer configuration is reduced when $\Delta$ is close to 0$^{\circ}$ or 180$^{\circ}$ -- keeping $\Delta$ close to 90$^{\circ}$ therefore maximises the accuracy of the measurements. \cite{Woollam_SE} A carefully defined modelling and fitting procedure is required to extract the energies corresponding to critical points in the band structure from the measured SE data. The semiconductor layers of the samples were described using a parametrised physical model which represents the sample dielectric constant as a sum of multiple Gaussian oscillators. \cite{Kim_PRB_1992} The presence of a native oxide layer on the surface of each sample was accounted for explicitly via tabulated dielectric constant data. \cite{Zollner_APL_1993} Using this model for each sample an iterative fitting procedure was then implemented to minimise the difference between the measured and simulated SE data, allowing the energies corresponding to the $E_{g}^{\Gamma}$ and $E_{g}^{\Gamma} + \Delta_{\scalebox{0.6}{\textrm{SO}}}$ interband transitions to be extracted.

The solid red, green and blue lines in Figs.~\ref{fig:ellipsometry_experiment}(a), (b) and (c) respectively show the measured SE data -- where $\tan( \Psi )$ is the amplitude ratio of the $p$- and $s$-polarisations after reflection \cite{Woollam_SE} -- for incident beam angles of 73.5$^{\circ}$, 74.0$^{\circ}$ and 74.5$^{\circ}$ in the GaP, GaP$_{0.987}$Bi$_{0.013}$ and GaP$_{0.963}$Bi$_{0.037}$ samples. Solid (dashed) black lines show the corresponding fits to $\Delta$ ($\Psi$) obtained using the procedure outlined above. The GaP sample consists of an epitaxial GaP buffer layer grown on a GaP substrate, and was analysed first in order to obtain accurate input parameters for the SE fits. These parameters were then used to describe the substrate and buffer layer in the subsequent models of the Bi-containing samples. Following this procedure it was possible to achieve good fits to the key features observed in the measured GaP$_{1-x}$Bi$_{x}$ SE spectra (cf.~Figs.~\ref{fig:ellipsometry_experiment}(b) and (c)). In Fig.~\ref{fig:ellipsometry_experiment}(a) a clear feature associated with $E_{g}^{\Gamma}$ is visible in the measured $\Delta$ and $\Psi$ spectra, which is well described by a modelling fit corresponding to a $\Gamma$-point GaP room temperature band gap $E_{g}^{\Gamma} = 2.76$ eV,  denoted in Fig.~\ref{fig:ellipsometry_experiment}(a) by a vertical arrow. The slight deviation from the accepted value of 2.78 eV is attributable to the sample growth taking place on n-doped GaP substrates. \cite{Jain_JAP_1990}

\begin{figure*}[t!]
     \centering
     \includegraphics[width=1.00\textwidth]{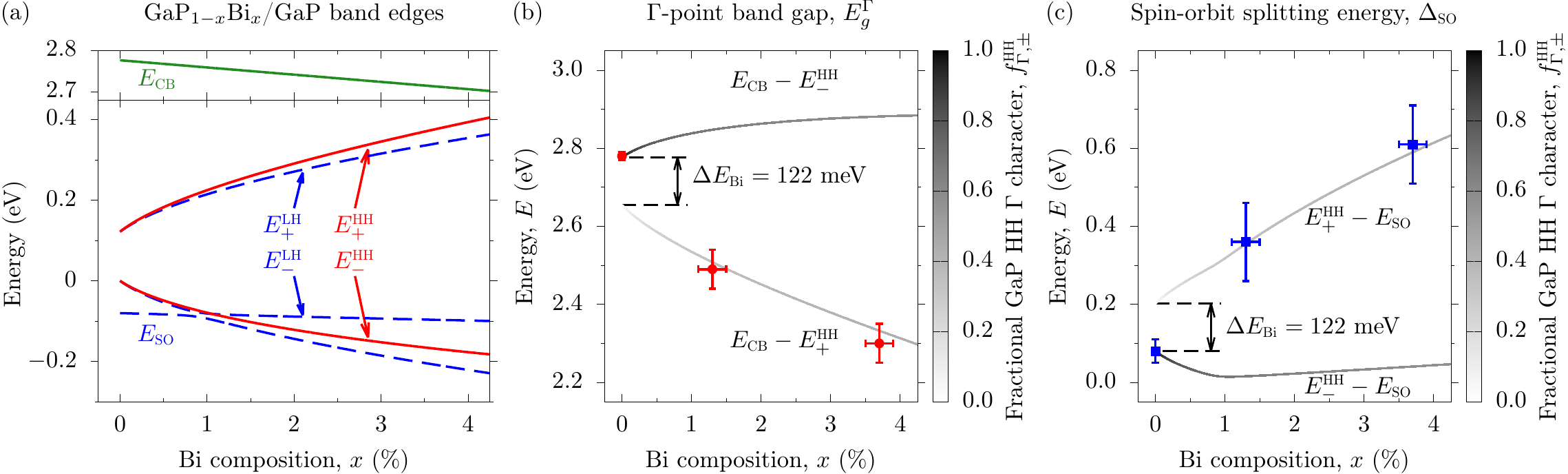}
     \caption{(a) Calculated variation of the $\Gamma$-point band edge energies with $x$ in pseudomorphically strained GaP$_{1-x}$Bi$_{x}$/GaP. Solid red, dashed blue and solid green lines respectively show the variation of the HH-like, LH/SO-like, and CB edge energies. (b) Variation of the GaP$_{1-x}$Bi$_{x}$/GaP $\Gamma$-point band gaps $E_{\protect\scalebox{0.6}{\textrm{CB}}} - E_{\pm}^{\protect\scalebox{0.6}{\textrm{HH}}}$ with $x$, calculated (shaded lines) and extracted from SE measurements (closed red circles). The line shading is determined by the fractional GaP HH $\Gamma$ character $f_{\Gamma,\pm}^{\protect\scalebox{0.6}{\textrm{HH}}}$ of the associated HH-like GaP$_{1-x}$Bi$_{x}$ VB states $E_{\pm}^{\protect\scalebox{0.6}{\textrm{HH}}}$. (c) Variation of the GaP$_{1-x}$Bi$_{x}$/GaP VB spin-orbit splitting energies $E_{\pm}^{\protect\scalebox{0.6}{\textrm{HH}}} - E_{\protect\scalebox{0.6}{\textrm{SO}}}$ with $x$, calculated (shaded lines) and extracted from SE measurements (closed blue circles). The line shading is as in (b).}
     \label{fig:vbac_theory}
\end{figure*}

Turning to Figs.~\ref{fig:ellipsometry_experiment}(b) and (c) it is clear from the measured $\Delta$ spectra that Bi incorporation gives rise to an additional feature which appears on the low energy side of the GaP direct band gap and shifts to lower energies with increasing $x$. This indicates a large reduction of $E_{g}^{\Gamma}$, in accordance with theoretical predictions. \cite{Usman_PRB_2011,Polak_SST_2015,Samadjar_MSSP_2015} The spectral features associated with $E_{g}^{\Gamma}$ are significantly broader in the Bi-containing samples than in GaP. This is likely associated with the presence of Bi composition fluctuations across the samples, as well as short-range alloy disorder. \cite{Usman_PRB_2013,Luo_NPGAM_2017} Using the multiple oscillator approach described above it was also possible to extract the energies associated with the $E_{g}^{\Gamma} + \Delta_{\scalebox{0.6}{\textrm{SO}}}$ transitions in each sample. The values of $E_{g}^{\Gamma}$ and $\Delta_{\scalebox{0.6}{\textrm{SO}}}$ extracted in this manner are shown respectively in Figs.~\ref{fig:vbac_theory}(b) and (c), using closed red circles and blue squares. We note that the uncertainties in these data are associated with the broadening of the associated features in the measured spectra. Overall, the SE measurements indicate that incorporation of dilute concentrations of Bi is sufficient to cause a giant reduction (increase) and bowing of $E_{g}^{\Gamma}$ ($\Delta_{\scalebox{0.6}{\textrm{SO}}}$).


To understand this unusual behaviour we have used supercell electronic structure calculations to analyse the contributions to the Bi-induced changes in the band edge energies, and to parametrise a suitable VBAC model for GaP$_{1-x}$Bi$_{x}$. This approach does not rely on post hoc fitting to alloy band structure data, thereby providing a predictive capability commonly lacking in models of this type. \cite{Broderick_SST_2013} In Ref.~\onlinecite{Usman_PRB_2011} we employed an atomistic tight-binding (TB) model to analyse the electronic structure of ordered and disordered GaP$_{1-x}$Bi$_{x}$ alloys. By directly constructing the $T_{2}$-symmetric localised states $\vert \psi_{\scalebox{0.6}{\textrm{Bi}}} \rangle$ associated with an isolated, substitutional Bi impurity we predicted the presence of a VBAC interaction having a composition dependence $\beta \sqrt{x}$. In the dilute doping (large supercell) limit we determined that the Bi-related localised states in GaP:Bi lie approximately 120 meV above the unperturbed GaP VB edge, in good agreement with experiment. \cite{Trumbore_APL_1966} Analysis of the electronic structure of ordered Ga(P,As)$_{1-x}$Bi$_{x}$ alloys eludicates the differences in the impact of Bi incorporation on the band structure: the natural VB offsets between GaP, GaAs and GaBi lead to the 6$p$ valence orbitals of Bi lying below the 4$p$ valence orbitals of As in energy, but higher in energy than the 3$p$ valence orbitals of P. As such, a substitutional Bi impurity forms a resonant localised state lying energetically below the VB edge in GaAs, but a bound localised state lying above the VB edge in energy in GaP. \cite{Usman_PRB_2011}

Building on our initial analysis of Ga(P,As)$_{1-x}$Bi$_{x}$ we have derived an extended basis set 12-band (VBAC) \textbf{k}$\cdot$\textbf{p} Hamiltonian to describe the dilute bismide band structure. \cite{Broderick_SST_2013} Using the TB model of Ref.~\onlinecite{Usman_PRB_2011} we have directly evaluated the Bi-related parameters of this model, including the distinct VBAC, virtual crystal (VC) and strain-related contributions to the Bi-induced shifts in the band edge energies. \cite{Broderick_SST_2015} To analyse the SE measurements we focus on the band edge energies at the zone centre: at $\Gamma$ the 12-band Hamiltonian diagonalises into decoupled blocks describing the CB, heavy-hole (HH), and light- and spin-split-off-hole (LH and SO) band edges. \cite{Broderick_SST_2015} As in GaAs$_{1-x}$Bi$_{x}$, the energy of the GaP$_{1-x}$Bi$_{x}$ $\Gamma$-point CB state $\Gamma_{6c}$ is well described as $E_{\scalebox{0.6}{\textrm{CB}}} (x) = E_{g}^{\Gamma} (0) - \alpha \, x + \delta E_{\scalebox{0.6}{\textrm{CB}}}^{\scalebox{0.6}{\textrm{hy}}}$, where the zero of energy has been chosen at the unperturbed GaP VB edge, $E_{g}^{\Gamma} (0) = 2.78$ eV is the host matrix band gap, $\alpha$ describes the VC shift of the CB edge energy, and $\delta E_{\scalebox{0.6}{\textrm{CB}}}^{\scalebox{0.6}{\textrm{hy}}}$ is the energy shift associated with the hydrostatic component of the compressive pseudomorphic strain in a GaP$_{1-x}$Bi$_{x}$/GaP epitaxial layer. \cite{Broderick_SST_2015}

The energies of the HH-like alloy VB states are given in the 12-band VBAC model as the eigenvalues of \cite{Usman_PRB_2011,Broderick_SST_2015}


\begin{equation}
     \left( \begin{array}{cc}
     \Delta E_{\scalebox{0.6}{\textrm{Bi}}} + \delta E_{\scalebox{0.6}{\textrm{Bi}}}^{\scalebox{0.6}{\textrm{hy}}} - \delta E_{\scalebox{0.6}{\textrm{Bi}}}^{\scalebox{0.6}{\textrm{ax}}} & \beta \sqrt{x} \\
     \beta \sqrt{x} & \kappa \, x + \delta E_{\scalebox{0.6}{\textrm{VB}}}^{\scalebox{0.6}{\textrm{hy}}} - \delta E_{\scalebox{0.6}{\textrm{VB}}}^{\scalebox{0.6}{\textrm{ax}}} \\ \end{array} \right) \begin{array}{c}
     \vert \psi_{\scalebox{0.6}{\textrm{Bi}}}^{\scalebox{0.6}{\textrm{HH}}} \rangle \\
     \vert \psi_{\scalebox{0.6}{\textrm{HH}}}^{\scalebox{0.6}{(0)}} \rangle \end{array} \, ,
     \label{eq:hh_hamiltonian}
\end{equation}

\noindent
where $\kappa \, x + \delta E_{\scalebox{0.6}{\textrm{VB}}}^{\scalebox{0.6}{\textrm{hy}}} - \delta E_{\scalebox{0.6}{\textrm{VB}}}^{\scalebox{0.6}{\textrm{ax}}}$ describes the VC, hydrostatic and axial strain-induced shifts to the GaP HH band edge energy, and $\Delta E_{\scalebox{0.6}{\textrm{Bi}}} + \delta E_{\scalebox{0.6}{\textrm{Bi}}}^{\scalebox{0.6}{\textrm{hy}}} - \delta E_{\scalebox{0.6}{\textrm{Bi}}}^{\scalebox{0.6}{\textrm{ax}}}$ is the energy of the HH-like Bi-related localised states relative to the zero of energy at the unperturbed GaP VB edge. \cite{Broderick_SST_2015} The energies of the LH- and SO-like VB states are given as the eigenvalues of a $3 \times 3$ matrix which can be found, along with full details of the model, in Ref.~\onlinecite{Broderick_SST_2015}.

The Bi-related band structure parameters computed for GaP$_{1-x}$Bi$_{x}$ are summarised in Table~\ref{tab:parameters}, where they are compared to those computed previously for GaAs$_{1-x}$Bi$_{x}$. The calculated VC parameters $\alpha$, $\kappa$ and $\gamma$ -- the latter describing the SO band edge energy \cite{Broderick_SST_2013,Broderick_SST_2015} -- are close to those calculated for GaAs$_{1-x}$Bi$_{x}$, reflecting (i) the larger (smaller) CB and VB (SO) offsets between GaP and GaBi than between GaAs and GaBi, \cite{Usman_PRB_2011} and (ii) the larger lattice mismatch between GaP and GaBi ($\approx 14$\%) than between GaAs and GaBi ($\approx 11$\%). The calculated VBAC coupling strength $\beta$ is larger in GaP$_{1-x}$Bi$_{x}$, reflecting the larger differences in size and electronegativity between P and Bi than between As and Bi. We note that $\beta$ in GaP$_{1-x}$Bi$_{x}$ is comparable to that calculated previously for the GaN$_{x}$P$_{1-x}$ CB ($\beta = 1.74$ eV). \cite{Harris_JPCM_2008,Reilly_SST_2009}


\begin{table}[t!]
	\caption{\label{tab:parameters} Bi-related parameters for the 12-band (VBAC) \textbf{k}$\cdot$\textbf{p} Hamiltonian of Ga(P,As)$_{1-x}$Bi$_{x}$, computed using atomistic TB calculations on ordered alloy supercells. The energy $\Delta E_{\protect\scalebox{0.6}{\textrm{Bi}}}$ of the Bi-related localised impurity states is given relative to the unperturbed Ga(P,As) host matrix VB edge.  \cite{Broderick_SST_2013,Broderick_SST_2015}}
	\begin{ruledtabular}
		\begin{tabular}{ccc}
			Parameter                                     & GaP$_{1-x}$Bi$_{x}$  & GaAs$_{1-x}$Bi$_{x}$ \\
			\hline
			$\Delta E_{\scalebox{0.6}{\textrm{Bi}}}$ (eV) & $0.122$              & $-0.183$ \\
			$\alpha$                                 (eV) & $~3.22$              & $~2.82$  \\
			$\beta$                                  (eV) & $~1.41$              & $~1.13$  \\
			$\gamma$                                 (eV) & $~0.24$              & $~0.55$  \\
			$\kappa$                                 (eV) & $~1.47$              & $~1.01$
		\end{tabular}
	\end{ruledtabular}
\end{table}



Figure~\ref{fig:vbac_theory}(a) shows the calculated variation of the $\Gamma$-point band edge energies with $x$ in pseudomorphically strained GaP$_{1-x}$Bi$_{x}$/GaP, for the CB ($E_{\scalebox{0.6}{\textrm{CB}}}$, solid green line), HH ($E_{\pm}^{\scalebox{0.6}{\textrm{HH}}}$, solid red lines), and LH and SO ($E_{\pm}^{\scalebox{0.6}{\textrm{LH}}}$ and $E_{\scalebox{0.6}{\textrm{SO}}}$, dashed blue lines) states, respectively. At $\Gamma$ the hydrostatic component of the pseudomorphic strain acts to push the CB (HH, LH and SO) edge(s) upwards (downwards) in energy, while the axial component lifts the degeneracy of VB edge to push HH-like states higher in energy than the LH-like states. In GaP$_{1-x}$Bi$_{x}$/GaP We calculate that $E_{\scalebox{0.6}{\textrm{CB}}}$ reduces linearly with increasing $x$, by 18 meV per \% Bi. The VBAC interaction produces two Bi-hybridised HH-like bands, the energies $E_{\pm}^{\scalebox{0.6}{\textrm{HH}}}$ of which vary strongly with $x$, displaying strong composition-dependent bowing. Beginning from $E_{-}^{\scalebox{0.6}{\textrm{HH}}} = 0$ and $E_{+}^{\scalebox{0.6}{\textrm{HH}}} = \Delta E_{\scalebox{0.6}{\textrm{Bi}}}$ at $x = 0$, we calculate that $E_{-}^{\scalebox{0.6}{\textrm{HH}}}$ ($E_{+}^{\scalebox{0.6}{\textrm{HH}}}$) decreases (increases) by 79 meV (103 meV) between $x = 0$ and 1\%. Similarly, the VBAC interaction produces a set of LH- and SO-like Bi-hybridised bands, \cite{Broderick_SST_2015} the energies of which are again strongly dependent on $x$ and characterised by strong composition-dependent bowing. As $E_{-}^{\scalebox{0.6}{\textrm{LH}}}$ moves downwards in energy towards $E_{\scalebox{0.6}{\textrm{SO}}}$ with increasing $x$ the coupling between the LH- and SO-like states -- brought about by the axial component of the pseudomorphic strain \cite{Broderick_SST_2015} -- leads to an anticrossing which is manifested in an abrupt increase in the rate at which $E_{\scalebox{0.6}{\textrm{SO}}}$ decreases for $x \gtrsim 1$\%. We note also that this axial strain-induced anticrossing between the LH- and SO-like states leads to a change of the VB ordering at $\Gamma$, with $E_{-}^{\scalebox{0.6}{\textrm{LH}}} > E_{-}^{\scalebox{0.6}{\textrm{HH}}}$ for $x \gtrsim 1$\%.


To compare the composition dependence of $E_{g}^{\Gamma}$ and $\Delta_{\scalebox{0.6}{\textrm{SO}}}$ inferred from the SE measurements to the theoretical calculations it is necessary to analyse the character of the VB eigenstates, in order to identify the presence of optically active states in the highly perturbed GaP$_{1-x}$Bi$_{x}$ VB. Since GaP$_{1-x}$Bi$_{x}$/GaP epitaxial layers are in a state of compressive pseudomorphic strain the highest energy alloy VB states are expected to be HH-like (cf.~Fig.~\ref{fig:vbac_theory}(a)). It is therefore sufficient to investigate the character of the eigenstates $\vert \psi_{\pm}^{\scalebox{0.6}{\textrm{HH}}} \rangle = a_{\scalebox{0.6}{\textrm{HH}}}^{\scalebox{0.6}{($\pm$)}} \, \vert \psi_{\scalebox{0.6}{\textrm{HH}}}^{\scalebox{0.6}{(0)}} \rangle + a_{\scalebox{0.6}{\textrm{Bi}}}^{\scalebox{0.6}{($\pm$)}} \, \vert \psi_{\scalebox{0.6}{\textrm{Bi}}}^{\scalebox{0.6}{\textrm{HH}}} \rangle$ of Eq.~\eqref{eq:hh_hamiltonian} -- corresponding respectively to the eigenvalues $E_{\pm}^{\scalebox{0.6}{\textrm{HH}}}$ -- which are a linear combination of the extended HH band edge state $\vert \psi_{\scalebox{0.6}{\textrm{HH}}}^{\scalebox{0.6}{(0)}} \rangle$ of the unperturbed GaP host matrix, and the HH-like Bi-related localised state $\vert \psi_{\scalebox{0.6}{\textrm{Bi}}}^{\scalebox{0.6}{\textrm{HH}}} \rangle$.

Since $\Delta E_{\scalebox{0.6}{\textrm{Bi}}} > 0$ in GaP$_{1-x}$Bi$_{x}$ the higher energy $E_{+}^{\scalebox{0.6}{\textrm{HH}}}$ eigenstate $\vert \psi_{+}^{\scalebox{0.6}{\textrm{HH}}} \rangle$ of Eq.~\eqref{eq:hh_hamiltonian} is primarily Bi-derived ($\vert a_{\scalebox{0.6}{\textrm{HH}}}^{\scalebox{0.6}{($+$)}} \vert^{2} < \frac{1}{2}$). Furthermore, given that the Bi-related localised states $\vert \psi_{\scalebox{0.6}{\textrm{Bi}}}^{\scalebox{0.6}{\textrm{HH}}} \rangle$ do not couple optically to the $\Gamma_{6c}$ CB edge states, \cite{Broderick_SST_2013} any optical transitions between $\vert \psi_{+}^{\scalebox{0.6}{\textrm{HH}}} \rangle$ and the $\Gamma$-point CB edge, having energy $E_{\scalebox{0.6}{\textrm{CB}}} - E_{+}^{\scalebox{0.6}{\textrm{HH}}}$, result from the VBAC interaction imparting GaP HH fractional $\Gamma$ character $f_{\Gamma,+}^{\scalebox{0.6}{\textrm{HH}}} \equiv \vert \langle \psi_{\scalebox{0.6}{\textrm{HH}}}^{(0)} \vert \psi_{+}^{\scalebox{0.6}{\textrm{HH}}} \rangle \vert^{2} = \vert a_{\scalebox{0.6}{\textrm{HH}}}^{\scalebox{0.6}{(+)}} \vert^{2}$ to $\vert \psi_{+}^{\scalebox{0.6}{\textrm{HH}}} \rangle$. Using Eq.~\eqref{eq:hh_hamiltonian}, $f_{\Gamma,+}^{\scalebox{0.6}{\textrm{HH}}}$ can be determined analytically as


\begin{equation}
     f_{\Gamma,+}^{\scalebox{0.6}{\textrm{HH}}} = \frac{ \beta^{2} x }{ \beta^{2} x + \left( E_{+}^{\scalebox{0.6}{\textrm{HH}}} - \kappa \, x - \delta E_{\scalebox{0.6}{\textrm{VB}}}^{\scalebox{0.6}{\textrm{hy}}} + \delta E_{\scalebox{0.6}{\textrm{VB}}}^{\scalebox{0.6}{\textrm{ax}}} \right)^{2} } \, .
     \label{eq:hh_character}
\end{equation}

As $x$ increases the increase in the strength $\beta \sqrt{x}$ of the VBAC interaction leads to $\vert \psi_{+}^{\scalebox{0.6}{\textrm{HH}}} \rangle$ acquiring significant GaP HH $\Gamma$ character which, despite being limited to values $< \frac{1}{2}$, is sufficient to produce appreciable optical coupling to the $\Gamma_{6c}$ CB states. For example, at $x = 1$\% we calculate $f_{\Gamma,+}^{\scalebox{0.6}{\textrm{HH}}} = 0.315$, indicating that the optical transition strength between $\vert \psi_{+}^{\scalebox{0.6}{\textrm{HH}}} \rangle$ and $\Gamma_{6c}$ in an \textit{ordered} GaP$_{0.99}$Bi$_{0.01}$ alloy should be close to one-third of that between $\Gamma_{8v}$ and $\Gamma_{6c}$ in GaP. Our analysis therefore suggests the emergence of an optically active band of primarily Bi-derived impurity states at energy $E_{+}^{\scalebox{0.6}{\textrm{HH}}}$.


In order to reflect this admixture of GaP Bloch and Bi localised character, we have calculated each of the potentially observable energy gaps $E_{g}^{\Gamma} = E_{\scalebox{0.6}{\textrm{CB}}} - E_{\pm}^{\scalebox{0.6}{\textrm{HH}}}$ and $\Delta_{\scalebox{0.6}{\textrm{SO}}} = E_{\pm}^{\scalebox{0.6}{\textrm{HH}}} - E_{\scalebox{0.6}{\textrm{SO}}}$. The results of these calculations are shown respectively in Figs.~\ref{fig:vbac_theory}(b) and~\ref{fig:vbac_theory}(c) for $E_{g}^{\Gamma}$ and $\Delta_{\scalebox{0.6}{\textrm{SO}}}$ (shaded lines), where they are compared to the measured values of $E_{g}^{\Gamma}$ and $\Delta_{\scalebox{0.6}{\textrm{SO}}}$ extracted from the SE measurements of Figs.~\ref{fig:ellipsometry_experiment}(a) --~(c). To describe the potential optical activity of these transitions the lines denoting the calculated transition energies are shaded according to the GaP HH $\Gamma$ character $f_{\Gamma,\pm}^{\scalebox{0.6}{\textrm{HH}}}$ of the corresponding HH-like alloy VB edge states $\vert \psi_{\pm}^{\scalebox{0.6}{\textrm{HH}}} \rangle$, with solid black describing a purely GaP-like state having $f_{\Gamma,-}^{\scalebox{0.6}{\textrm{HH}}} \equiv \vert a_{\scalebox{0.6}{\textrm{HH}}}^{(-)} \vert^{2} = 1$. No clear features were distinguishable in the measured SE spectra close to the calculated transition energies $E_{\scalebox{0.6}{\textrm{CB}}} - E_{-}^{\scalebox{0.6}{\textrm{HH}}}$, while the calculated transition energies $E_{-}^{\scalebox{0.6}{\textrm{HH}}} - E_{\scalebox{0.6}{\textrm{SO}}}$ lie outside the spectral range of the experiment.

The quantitative agreement between the calculated and measured data in Figs.~\ref{fig:vbac_theory}(b) and (c) confirms that the extremely large observed reduction (increase) and bowing of $E_{g}^{\Gamma}$ ($\Delta_{\scalebox{0.6}{\textrm{SO}}}$) results from the emergence of an optically active band of primarily Bi-derived impurity states lying energetically within the GaP band gap. That this impurity band lies energetically within the host matrix band gap accounts quantitatively for the observed trends: the contribution of the strong, composition-dependent bowing of $E_{+}^{\scalebox{0.6}{\textrm{HH}}}$ to the decrease (increase) of $E_{g}^{\Gamma}$ ($\Delta_{\scalebox{0.6}{\textrm{SO}}}$) is combined with the binding energy $\Delta E_{\scalebox{0.6}{\textrm{Bi}}}$ of the Bi-related localised states. This behaviour is qualitatively distinct from that in GaAs$_{1-x}$Bi$_{x}$, where substitutional Bi impurities generate localised states which are resonant with the GaAs VB, \cite{Zhang_PRB_2005,Usman_PRB_2011,Joshya_PRB_2014,Alberi_PRB_2015} but similar to that in the dilute nitride alloy GaN$_{x}$P$_{1-x}$, where substitutional N impurities produce a band of primarily N-derived states lying deep within the GaP band gap, several hundred meV below the $\Gamma_{6c}$ CB state. \cite{Shan_APL_2000,Zhang_PRB_2000,Kent_PRB_2001,Wu_PRB_2002,Fluegel_PRB_2005,Gungerick_PRB_2006,Harris_JPCM_2008,Reilly_SST_2009}

From the SE measurements we extract $E_{g}^{\Gamma} = 2.49$ eV at $x = 1.3$\% ($f_{\Gamma,+}^{\scalebox{0.6}{\textrm{HH}}} = 0.338$), an extremely large reduction of 270 meV compared to the measured GaP $\Gamma$-point band gap $E_{g}^{\Gamma} (0) = 2.78$ eV. This is in exact agreement with the calculated reduction of 270 meV in $E_{g}^{\Gamma}$ between $x = 0$ and 1.3\% in pseudmorphically strained GaP$_{1-x}$Bi$_{x}$/GaP. Given the calculated reduction of 23 meV in $E_{\scalebox{0.6}{\textrm{CB}}}$ between $x = 0$ and 1.3\%, we conclude that the majority (91\%) of the reduction in $E_{g}^{\Gamma}$ is associated with the emergence of the $E_{+}^{\scalebox{0.6}{\textrm{HH}}}$ impurity band. Similarly, we measure an extremely large ($>$ fourfold) increase of $\Delta_{\scalebox{0.6}{\textrm{SO}}}$, from 80 meV in GaP to approximately 360 meV at $x = 1.3$\%. This is again in excellent agreement with the calculated value $\Delta_{\scalebox{0.6}{\textrm{SO}}} = 357$ meV, with the majority (92\%) of the increase in $\Delta_{\scalebox{0.6}{\textrm{SO}}}$ associated with the emergence of the $E_{+}^{\scalebox{0.6}{\textrm{HH}}}$ band.

Increasing the Bi composition from 1.3 to 3.7\% we note that the relative change in $E_{g}^{\Gamma}$ and $\Delta_{\scalebox{0.6}{\textrm{SO}}}$ per \% Bi is significantly reduced. The measured (calculated) value $E_{g}^{\Gamma} = 2.30$ eV (2.33 eV) at $x = 3.7$\% ($f_{\Gamma,+}^{\scalebox{0.6}{\textrm{HH}}} = 0.421$) represents a further reduction of 190 meV (175 meV) from $x = 1.3$\%, while the measured (calculated) value $\Delta_{\scalebox{0.6}{\textrm{SO}}} = 0.61$ eV (0.59 eV) at $x = 3.7$\% Bi represents a further increase of 250 meV (230 meV) from $x = 1.3$\%. For $E_{g}^{\Gamma}$ this change is only 86\% of that between $x = 0$ and 1.3\%, despite occuring over a 2.4\% increase in $x$. The measured and calculated changes of $\Delta_{\scalebox{0.6}{\textrm{SO}}}$ between $x = 1.3$ and 3.7\% are approximately equal to those between $x = 0$ and 1.3\%, again representing a significantly reduced change per \% Bi. These trends highlight the strong dependence of the bowing of $E_{g}^{\Gamma}$ and $\Delta_{\scalebox{0.6}{\textrm{SO}}}$ on $x$. Our measured and calculated variation of $E_{g}^{\Gamma}$ and $\Delta_{\scalebox{0.6}{\textrm{SO}}}$ with $x$ differs significantly from those predicted using first principles calculations, \cite{Polak_SST_2015} but are close to those calculated via a VBAC model using parameter estimates extracted based on the available data for related alloys. \cite{Samadjar_MSSP_2015}

We now turn our attention to two key qualitative features of the GaP$_{1-x}$Bi$_{x}$ electronic structure. Firstly, GaP has an indirect band gap due to the X$_{6c}$ CB states lying $\approx 0.5$ eV below $\Gamma_{6c}$, while in semimetallic GaBi the X$_{6c}$ states lie $\approx 2$ eV below $\Gamma_{6c}$. \cite{Janotti_PRB_2002,Usman_PRB_2011} Applying the VC approximation in conjunction with the TB model we estimate that the X$_{6c}$ states shift downwards in energy by $\approx$ 12 meV per \% Bi in free-standing GaP$_{1-x}$Bi$_{x}$. This is less than the 32 meV per \% Bi reduction of the $\Gamma_{6c}$ state energy described by the VC parameter $\alpha$ (cf.~Table~\ref{tab:parameters}), suggesting that Bi incorporation may bring about a direct $\Gamma_{8v}$-$\Gamma_{6c}$ band gap for sufficiently high $x$. However, while our analysis predicts a reduction of the $X_{6c}$-$\Gamma_{6c}$ energy gap up to $x \approx 15$\%, for higher $x$ it indicates a step change in the CB structure: the $p$-like $\Gamma_{7,8c}$ states move lower in energy that the $s$-like $\Gamma_{6c}$ states, drastically increasing the $X_{6c}$-$\Gamma_{6c}$ energy gap. Our analysis therefore suggests that GaP$_{1-x}$Bi$_{x}$ remains an indirect-gap alloy, in agreement with the first principles calculations of Ref.~\onlinecite{Polak_SST_2015}. This highlights an important qualitative difference between the GaP$_{1-x}$Bi$_{x}$ and GaN$_{x}$P$_{1-x}$ band structures, since substitutional N in GaP generates localised states lying below the host matrix X$_{6c}$ states in energy, bringing about a quasi-direct band gap. \cite{Shan_APL_2000,Zhang_PRB_2000,Kent_PRB_2001,Wu_PRB_2002,Fluegel_PRB_2005,Harris_JPCM_2008}

Secondly, our analysis in Ref.~\onlinecite{Usman_PRB_2011} demonstrated that the VBAC description of the GaP$_{1-x}$Bi$_{x}$ VB structure breaks down rapidly with increasing $x$ in the presence of short-range alloy disorder. Bi clustering creates a distribution of Bi-related localised states -- lying across a range of energies within the GaP band gap -- with which the GaP VB edge states strongly hybridise. This leads to a distribution of GaP VB edge $\Gamma$ character over a multiplicity of impurity levels, suggesting that there is no single band that possesses sufficient Bloch character to allow for appreciable absorption or emission of light. While our results above demonstrate that the VBAC model provides a useful approach to analyse the main features of the band structure, the details of the electronic structure are in practice determined primarly by the impact of short-range alloy disorder. The high crystalline quality and difficulty in obtaining photoluminescence from the samples studied here \cite{Nattermann_JCG_2017_1} supports this interpretation: the GaP$_{1-x}$Bi$_{x}$ optical properties are intrinsically limited not solely by defects associated with the low growth temperatures required to incorporate Bi, but by a combination of an indirect band gap and a breakdown in VB edge Bloch character.

Despite having lattice constants commensurate with growth on Si, our analysis suggests that refinement of the epitaxial growth of GaP$_{1-x}$Bi$_{x}$ alloys is unlikely to lead to efficient light emitters: the optical properties are expected to be intrinsically limited by the nature of the material band structure. However, just as quaternary GaN$_{x}$As$_{y}$P$_{1-x-y}$ alloys have found applications in III-V semiconductor lasers monolithically integrated on Si, \cite{Liebich_APL_2011} it is possible that similar progress could be made using As-rich quaternary GaP$_{1-x-y}$As$_{y}$Bi$_{x}$ alloys for, e.g., applications in multi-junction solar cells due to the fact that they can be grown lattice-matched to either GaAs or germanium (Ge) while having band gaps close to 1 eV. \cite{Forghani_APL_2014,Nattermann_AMT_2016,Nattermann_JCG_2017_2}


In conclusion, we have presented a combined experimental and theoretical investigation of the GaP$_{1-x}$Bi$_{x}$ band structure. Measurements performed on GaP$_{1-x}$Bi$_{x}$/GaP epitaxial layers reveal giant bowing of $E_{g}^{\Gamma}$ and $\Delta_{\scalebox{0.6}{\textrm{SO}}}$, whereby $E_{g}^{\Gamma}$ ($\Delta_{\scalebox{0.6}{\textrm{SO}}}$) decreases (increases) by approximately 200 meV (240 meV) between $x = 0$ and 1\%. These changes are characterised by strong, composition-dependent bowing. Electronic structure calculations confirm that substitutional Bi in GaP generates localised impurity states lying energetically within the GaP band gap, and that the main features of the GaP$_{1-x}$Bi$_{x}$ band structure can be understood in terms of a VBAC interaction between the extended states of the GaP VB edge, and highly localised Bi-related impurity states. A VBAC model was derived and parametrised directly from atomistic supercell calculations, allowing quantitative prediction of the evolution of the main features of the band structure with $x$. Our analysis suggests that the highest energy VB in GaP$_{1-x}$Bi$_{x}$ is a hybridised impurity band: admixture of the GaP VB edge $\Gamma$ character into this primarily Bi-derived band allows optical coupling to the comparatively unperturbed CB states. Aspects of the GaP$_{1-x}$Bi$_{x}$ band structure are broadly comparable to GaN$_{x}$P$_{1-x}$, but key qualitative differences highlight the distinction between Bi and N as isovalent impurities in conventional III-V semiconductors.


This work was supported by the Engineering and Physical Sciences Research Council, U.K. (EPSRC; project nos.~EP/H005587/1, EP/N021037/1, and EP/K029665/1), by Science Foundation Ireland (SFI; project no.~15/IA/3082), and by the German Science Foundation (DFG; project no.~GRK 1782). Z.L.B.~acknowledges support from the University of Surrey Marion Redfearn and Advanced Technology Institute Scholarships. The data associated with this work are available from the University of Surrey publications repository at \url{http://epubs.surrey.ac.uk/XXXXXX/}.



\end{document}